\documentclass[paper]{JHEP}
\usepackage{epsf}
\usepackage{amsmath}
\usepackage{amsfonts}
\usepackage{amssymb}

\newcommand{\pa}{\partial}
\newcommand{\tr}{{\rm tr}}
\newcommand{\comment}[1]{}

\newcommand{\pasl}{\pa\kern-.55em /}

\newcommand{\ksl}{k\kern-.55em /}

\DeclareFixedFont{\xiiss}{OT1}{cmss}{m}{n}{12}
\DeclareFixedFont{\ixss}{OT1}{cmss}{m}{n}{9}
\DeclareFixedFont{\cmrnine}{OT1}{cmr}{m}{n}{9}
\newcommand{\field}[1]{\mathbb{#1}}
\newcommand{\BC}{{\field C}}

\newcommand{\BZ}{{\field Z}}

\newcommand{\CCs}{\hbox{\ixss C\kern-.4emI}}
\newcommand{\ZZs}{\hbox{\ixss Z\kern-.4emZ}}

\newcommand{\CA}{{\cal A}}

\newcommand{\CM}{{\cal M}}
\newcommand{\CZ}{{\cal Z}}

\newcommand{\ZA}{\CZ\CA}

\newcommand{\CP}{{\BC\field P}}

\newcommand{\diag}{\hbox{diag}}

 \newcommand{\myfig}[3]{\begin{figure}[ht]
\begin{center}
\leavevmode
\epsfxsize=#2cm
\epsfbox{#1}
\end{center}
\caption{#3}
\label{fig:#1}
\end{figure}}

\newcommand{\Ext}{{\cal E}xt}

\preprint{ILL-(TH)-01-04\\ hep-th/0105229}
\title{Resolution of Stringy Singularities by Non-commutative Algebras}

\author{David Berenstein\thanks{\email{berenste@pobox.hep.uiuc.edu}}
and Robert G. Leigh\thanks{\email{rgleigh@uiuc.edu}}
\\
Department of Physics\\
University of Illinois at Urbana-Champaign\\
Urbana, IL 61801}

\abstract{In this paper we propose a unified approach to 
(topological) string theory on certain singular spaces in their large
volume limit. The approach exploits the non-commutative structure of
D-branes, so the space is described by an algebraic geometry of
non-commutative rings. The paper is devoted to the study of 
examples of these algebras.
In our study there is an auxiliary commutative algebraic geometry
of the center of the (local)
algebras which plays an important role as the target space geometry
 where closed strings
propagate. The singularities that are resolved will be the 
singularities of this auxiliary geometry. The 
singularities are resolved by
the non-commutative algebra if the local non-commutative rings are
regular. This definition guarantees that D-branes 
have a well defined K-theory class. 
Homological functors also play an important role.
They  describe the intersection theory of D-branes and lead to a
formal definition of local quivers at singularities, which can be
computed explicitly for many types of singularities. 
These results can be interpreted in terms of the
derived category of coherent sheaves over the
non-commutative rings, giving a non-commutative version of recent work by
M. Douglas.
We also describe global features like the Betti
numbers of compact singular Calabi-Yau threefolds via global holomorphic
sections of cyclic homology classes. }


\begin{document}

\section{Introduction}

D-branes \cite{DLP,P} have played a pivotal role in changing our
understanding of string theory. Even though D-branes are nonperturbative
effects in string
theory, they are remarkably simple mainly because they admit a
description via
boundary conformal field theory. A D-brane configuration defines in some
sense
an open string field theory, and in open string theories
one can derive the closed string states from the knowledge of the open
string
amplitudes by analyzing the pole structure in the one loop amplitudes.

It has been known for a while that open string theories are inherently
non-commutative objects \cite{W}, that is, the operation of  gluing two
open strings distinguishes the ends of the strings at which they are
joined. It was not until the advent of Matrix theory \cite{BFSS} that
this important feature of open strings was appreciated, as there it was
seen that the target space coordinates  of D-branes are described by
matrices rather than just numbers. More recently in \cite{CDS} and
\cite{SW} it has been argued that field theories on non-commutative
spaces are  low energy limits of open string theories and the origin of
non-commutativity has been related to the NS B-field. As well, the
classification of topological D-brane charges is given by the
topological K-theory of the target space \cite{MM,Sen,WittenK}, which is
in fact a well defined group for any non-commutative algebra\footnote{In
string theory one can argue that the algebras in questions are $\BC^*$
algebras \cite{W}, so it is appropriate to consider the K-theory of a
$\BC^*$ algebra.} (see \cite{Blackadar} for example).

More recently we have found that non-commutative algebras give a
description of moduli spaces of D-branes on certain orbifold
singularities \cite{BJL}, and that this description is sufficient to
calculate the Betti numbers of certain singular orbifold Calabi-Yau
spaces \cite{BL4} without computing the topological closed string theory
spectrum.

There are compactification spaces of string theory which are singular in
the large volume limit as commutative geometries (orbifolds with fixed
points for example), whereas the string theory on these spaces is
perfectly smooth \cite{DHVW,DHVW2} since one can compute all the
correlation functions in the CFT. If one resolves all of the
singularities geometrically (either by blow-ups or deformations, see
\cite{AGM,AGM2,AMG} for example) then one can calculate the stringy data
for these deformed spaces and take the limit towards the singular
geometry to describe the singular space. There are exceptions to this
argument: not all singularities can be resolved in this manner, as was
shown for a particular example of orbifolds with discrete torsion
\cite{VW}. There the generic deformation of the Calabi Yau space has
conifold singularities which cannot be blown up or deformed away.
However, in \cite{BL4} we showed that these spaces admit an elegant
non-commutative description. If one considers matrix theory as a
prototype, it is natural to define the target space geometry as that
which is seen by D-branes, and ask questions about closed strings later.
That is, instead of defining a target space as a closed string theory
background, one would prefer to ask what are the allowed configurations
of D-branes and infer the closed string background from this data.
Naturally, D-branes via their boundary states can carry all of the
possible closed string state information, as they are sources for all
possible closed string states. In some sense, the full D-brane data can
be as complicated as the closed string theory of the background itself, 
so we opt for a less detailed description in terms of topological 
low energy degrees of freedom alone.

Our intention in this paper is to make a general proposal for
`resolving' singularities within non-commutative geometry and to
understand the D-branes on these spaces. The main idea is that even when
one has blown down a cycle to obtain a singularity, one can still have a
$B$ field through the shrunken cycle. Geometrically we have
\begin{equation}
\frac {\int_\Sigma Re (\omega)}{{\int_\Sigma Im (\omega)}}\to \infty
\end{equation}
(where $\omega=B+iJ$ is the complexified K\"ahler form)
so the $B$ field is locally very large in geometric units. From the
arguments of Ref. \cite{SW} this should signal that the geometry of the
singularity is well described
by a non-commutative geometry.

The basic outline of our program is as follows: one is given a
non-commutative complex manifold, that is, a collection of holomorphic
non-commutative algebras describing analytic coordinate patches on the
complex manifold, and maps on how to glue these algebras. In our case
the algebras on each patch $\CA_i$ have a center $\ZA_i$, and the center
is glued to give a commutative algebra. The algebraic geometry of this
commutative algebra will be identified with the target space where
closed strings propagate, which might be singular, and the algebraic
geometry of the non-commutative algebras $\CA_i$ will give the
resolution of those singularities. This is the spirit of \cite{SW} where
there are two geometries at play, one for closed strings and another one
for open strings. In this sense a commutative singularity can be made
smooth in a non-commutative sense. We are not making the claim however
that {\it any} singularity can be resolved in this way; those that admit
conformal field theory descriptions should admit this type of
resolution. The notion of smoothness will be given by an algebraic
definition (namely that locally the algebras be regular), and it is
intimately tied up to the derived category of coherent sheaves of the
algebra (that is, the D-branes are giving us a definition of
smoothness). Here we find the approach of \cite{Kont,Doucat} very
illuminating. One should be able to define the target space of string
theory by using the derived category of coherent sheaves (holomorphic
$B$ model D-branes), which are the topological branes. This approach has
natural advantages. Once we satisfy our definition of smoothness every
coherent sheaf will admit a (locally) projective resolution. That is,
any brane can be obtained from branes and antibranes on the
non-commutative manifold which look like vector bundles, with a choice
of tachyon profile. This is desirable because it tells us that the
classification of D-branes is given by the algebraic K-theory of the
non-commutative space \cite{Ros}. For orbifolds of a smooth space, the
construction reproduces the quiver diagrams for brane fractionation (\`a
la Douglas-Moore \cite{DM}), and moreover the K-theory of this algebra
is automatically the (twisted) equivariant K-theory of the original
commutative algebra with respect to the orbifold group action (see
\cite{Connes,Blackadar} for example). The projective resolution also
allows us to compute the number of strings between two branes, and from
this data one can define the intersection number of pairs of branes.

In this paper we will only ask questions that do not depend on the
target space metric and instead work with just configurations of BPS
branes, which are equally well described by the topologically twisted
string theories associated to a target space. This is also the natural
setting for studying mirror symmetry and quantum cohomology
\cite{CoxKatz}, which make this coarser description a good place to find
everything which is calculable at a generic point in moduli space.

We have chosen to present our proposal through examples, and we will
deal with these as if they are generic rather than proceed with an
abstract theory. Our purpose is to find a constructive approach to these
algebras (the ones that resolve singularities) and how to glue them, as
well as to how one extracts computable data from the algebras.  In the
end this approach gives us our target space and it's Betti numbers, as
well as the category of D-branes on the space. A full description of the
homological and algebraic tools used in the paper may be found in Refs.
\cite{Ros,Wegge,Larsen,Blackadar,Connes,Loday,GManin,Weibel,RaeWil,Hart}.

The paper is organized as follows: In section \ref{sec:prelim} we begin
with an introduction to non-commutative algebraic geometry in the
setting of \cite{BJL}. In particular we define what a non-commutative
point is (a point-like brane)  and how to build the target space
geometry from these objects.

In Section \ref{sec:crossed} we present the principal algebraic
construction we study, and define the {\it crossed product} of an
algebra (which can be taken to be commutative) with a discrete group of
automorphisms. This is how one constructs an orbifold space as a
non-commutative geometry. We consider in the following sections a number
of examples, primarily the well-known $\BC^2/\BZ_n$ orbifold and
associated orbifolds with discrete torsion. Using the crossed product we
show that unorbifolding an Abelian orbifold gives an algebra which is
(locally) Morita equivalent to the original commutative algebra. In
particular this means that the K-theory of both algebras is the same,
but more is true, their D-brane categories are the same as well; so both
algebras are describing the same geometry.

In Section \ref{sec:orborb} we exploit ideas presented in \cite{BJL3} to
understand an orbifold of an orbifold; that is, we consider the problem
of orbifolding a general non-commutative space, and in particular we
focus on the problem of how one distinguishes ungauged symmetries from
gauged symmetries. This is not an issue for a commutative algebra, but
it is a subtle technical point which might provide anomalies to
construct backgrounds with certain geometric data in string theory, as
given  for example in \cite{Joyce}. 

In Section \ref{sec:conif}, we consider other non-orbifold
singularities, such as the conifold geometry and it's brane realization
\cite{KW}. We discuss what algebra should be associated with the
conifold. As well we recall the conifold of \cite{VW} and it's D-brane
realization \cite{D}. This demonstrates that two distinct
non-commutative geometries resolve the same geometric singularity, and
they are not Morita equivalent to each other, as the spectrum of
fractional branes at the singularity is different. This demonstrates
that these two conifolds are topologically distinct.

Next, in Section \ref{sec:inter} we consider the problem of calculating
the massless open string spectrum between two distinct D-branes, based
on ideas of Douglas \cite{Doucat} generalized to non-commutative
geometry. Here we find our definition of regularity: namely that any
D-brane locally can be associated to the decay product of a finite
number of (locally) space-filling branes and anti-branes (that is, any
coherent sheaf admits locally a projective resolution). The number of
strings between two branes is topological and in this setting
corresponds to calculating the derived functor for the module morphisms
$hom(A,B)$, namely the $\Ext$ groups. Here is where we first find a
computation in homological algebra and where we see a clear connection
to \cite{Kont, Doucat}. In particular, the regularity condition
guarantees that the process of calculating these intersection numbers
terminates. We show in the example $\BC^2/\BZ_n$ how the quiver diagram
construction of \cite{DM} is exactly the computation of the groups
$\Ext^1(A,B)$ for $A,B$ fractional branes (which are the nodes of the
quiver diagram).

Following this construction, we focus our attention on closed strings in
Section \ref{sec:betti}. We argue that the proper tool to understand
topological closed string states is the cyclic homology of the algebra.
The reasons behind this proposal rest on the map between closed string
states and gauge invariant operators in the AdS/CFT correspondence,
which identify closed string states as traces. Cyclic homology is a
homology theory of generalized traces and it is an invariant under
Morita equivalence. Cyclic homology is dual to K-theory, as one can
produce a pairing between them. Thus it is the natural receptacle for
closed string states as it allows us to compute topological couplings
between D-branes and cohomology classes (RR topological fields). From
here we use our results to compute the Betti numbers of the orbifold
$T^6/\BZ_2\times\BZ_2$ with and without discrete torsion, and we show
how these are related to the topology of the fibration of the
non-commutative geometry over the singular set. Finally we conclude with
open problems and possible future lines of development.

\section{Algebraic preliminaries}\label{sec:prelim}

D-brane configurations may be given by systems of equations, which may
be thought of as the moduli space problem in the low energy field theory
limit. More generally, if one considers collections of D-branes and
anti-D-branes, one still gets a system of equations which 
are derived from the effective action of the tachyon condensate plus the massless
fields (everything else can be integrated out). 
In our approach these tachyons will be implicit, and will appear
later when we study the homological functors.
Our approach applies to topological branes, in
the sense that we are ignoring the metric (and also the B-field which is just
the complexification of the metric). 
The latter is important for calculating
Chern classes, which determine cohomology rather than 
K-theory classes, but
may otherwise be ignored. Thus, we can focus on holomorphic data, and
deal with the algebraic geometry of non-commutative algebras.

Our approach is based on geometric realizations of string theories.
Non-geometric phases of string theory are not described in this
framework. It is also a low energy approach, and assumes that  the
non-commutative algebras we begin with are a good description of the
space. With this philosophy in mind we sacrifice exact closed string
backgrounds such as those built out of RCFTs as well as their boundary
states (see \cite{DiaD,BDLR,RS,BDLR} for very interesting work in those
directions). However, our construction can describe generic points in
the moduli space of complex structures, for which no known exact
conformal field theory description exists. A good review of the ideas
necessary to understand D-branes at generic points can be found in
\cite{Douglas}. Further, we restrict ourselves to special classes of
algebras, which in some sense are almost commutative, which is exactly
the opposite limit that one considers when one studies deformation
quantization (as in the work \cite{ARS,SchomerusDQ} ). In principle our
results should admit generalizations in this direction, but the
algebraic tools will probably change drastically. The algebras that we
consider here  are finitely generated\footnote{ It is probably better to
work with the classes of algebras which are locally Morita equivalent to
one of these, but this requires a more elaborate structure
\cite{BM,RaeWil}.} over their (local) center; that is, we require the
non-commutative algebra to be a sheaf of algebras over a possibly
singular commmutative space. As we will mainly be interested in singular
examples to understand how the non-commutative geometry resolves the
space, we will assume that we are in a singular manifold already.

This choice of algebras seems rather restrictive. However, any global
orbifold space by a finite group action admits a description in terms of
these algebras, see section \ref{sec:crossed}, and thus any space which
has local singularities of the orbifold type also admits this kind of
description. In general it is probably useful to consider a more general
space defined by an aritrary $\BC^*$ algebra as a target space for
D-branes. However we might lose the commutative algebraic geometry tools
that are so useful when dealing with computations in Calabi Yau spaces.
Some of the tools we have described  carry over, but the tools required
to make these spaces tractable will look very different than the
presentation we are giving here \cite{Connes}, and moreover the
holomorphic structure of the space might be lost.

The above description we have given is essentially an Azumaya algebra
over the singular complex manifold, and these have already appeared in
\cite{Kap, Kap2} as describing branes on quotient spaces with some
B-field torsion, as well as branes on tori. However, our description is
local and not based on global quotients.

A simple example is that of $D$-branes at an orbifold $\BC^3/\BZ_n\times
\BZ_n$ with discrete torsion \cite{D,DF}. Here we get a $U(N)$ gauge
theory with three adjoints $\phi_{1,2,3}$ and superpotential
\begin{equation}
W = \tr(\phi_1\phi_2\phi_3- q\phi_2\phi_1\phi_3)
\end{equation}
with $q^n=1$.
The F-term constraints are then seen to be given by
\begin{equation}
\phi_1\phi_2- q \phi_2\phi_1 = 0 \label{eq:qcomm}
\end{equation}
and it's cyclic permutations, which are holomorphic constraints giving
rise to a non-commutative algebra.

The above algebra is a non-commutative algebra which encodes the target
space. In \cite{BJL} this theory was analyzed using non-commutative
geometry tools. Points in the (non-commutative) space were defined as
irreducible representations of the above algebra. By Schurs lemma, any
element of the center evaluated on a representation will be proportional
to the identity, and this value will give a point in the commutative
space of the center of the algebra. The center of the algebra thus gives
exactly the ring of holomorphic functions of the orbifold, what one
would naturally call the orbifold space: the target space for the closed
strings. Thus there are two geometries: one non-commutative associated
to the point-like D-branes, and a commutative one associated to the
closed strings, in a spirit much like the work of Seiberg and
Witten\cite{SW}. The non-commutative geometry is fibered over the
commutative space, and most of the geometric calculations we will do, as
well as our intuition, depend crucially on understanding this fibration
in detail. More details as to how this construction should be introduced
can be found in \cite{BJL}.

A generic D-brane will be a coherent sheaf of modules over the local
algebras which are finitely presented, and as such we are naturally in
the topological B-model. We will assume that we are in a large volume
phase of the Calabi Yau manifold (except for the singularities), so that
we are in a geometric phase of the string field theory (and not a Landau
Ginzburg phase for example), but in general we will ignore the issue of
the metric all together and we will just be interested in the questions
that depend only on the holomorphic structure of the space.

On the algebra defined by the relations \ref{eq:qcomm} it is natural to
construct the closed string spectrum associated to the singularity by
using the AdS/CFT correspondence. The one particle closed string states
are then the single trace elements of the chiral ring of the above
conformal field theory. The support of a closed string state is the set
of non-commutative points where evaluation of the trace gives a non-zero
result. It was found in \cite{BJL} that this chiral ring was precisely
the Hoschild homology in degree zero of the above algebra $HH_0(\CA)$.
This in turn is the same group as the cyclic homology group $HC_0(\CA)$,
see \cite{Loday} for more mathematical details, and it is suggestive to
build a theory of closed string states by using these groups.

The data given in general will be the local (on some patch) descriptions
of the non-commutative algebras as a set of generators and relations.
Given these, we need to calculate the center of the (local) algebra and
its representation theory. This procedure will give us a model for our
(local) target space. The prescription needs to be supplemented by a
recipe for gluing algebraic data (localization theory) and extracting
global invariants (global holomorphic sections, betti numbers, moduli
spaces, etc) of the space. The localization takes place by taking
denominators in the center of the algebra, namely, we glue the
`geometry' of the closed strings. For the global invariants we will only
develop the intersection theory of branes and the calculation of betti
numbers on orbifold Calabi Yau spaces.

\section{Crossed products}\label{sec:crossed}

The low energy degrees of freedom on supersymmetric point-like branes at
orbifold singularities are constructed by taking images of branes and
projecting onto gauge invariant states. This construction was carried
out in detail in \cite{DM} and gives rise to field theories which are
represented by a quiver diagram.

The construction exploits the fact that the group action on the orbifold
acts via a gauge transformation on the brane, and the low energy degrees
of freedom are encoded in the projection condition
\begin{equation}
\gamma(g) \phi^i \gamma(g)^{-1} = R^i_j \phi^j \label{eq:gaugeorb}
\end{equation}
where the $\phi^i$ are directions normal to the singularity, and $R^i_j$
is the representation under which they transform. The $\gamma(g)$ are
given by the regular representation of the group (possibly twisted by
some cocycle to account for discrete torsion \cite{D, DF}). The low
energy degrees of freedom are captured by the $\phi^i$ which satisfy the
above conditions, and the superpotential is derived from the
superpotential of the parent $N=4$ theory.

We want to read \ref{eq:gaugeorb} as an algebraic identity in the
non-commuttaive space. First, we have the $\phi^i$ which commute among
themselves (since they come from the unorbifolded theory whose F-terms
are $[\phi^i, \phi^j]=0$); these form an algebra $\CA\sim\BC^3$. Next,
according to (\ref{eq:gaugeorb}), we need to add generators to the
algebra for each $g$, which we will call $e_g$. These are such that
$e_g\cdot e_{g'} = \epsilon(g,g') e_{gg'}$ with $\epsilon$ the twisting
cocyle. The twisting cocycle is how one introduces discrete
torsion\cite{D,DF}.

In this algebra the outer automorphism of the commutative algebra
$[\phi^i, \phi^j]= 0$ given by the orbifold group action is gauged, and
becomes an inner automorphism of the non-commutative algebra, that is,
conjugation by the elements of $G$ becomes part of the algebra. This new
larger algebra is called the {\it crossed product algebra} (see
\cite{Blackadar} for details) of $\CA$ and $\Gamma G$ (the group
algebra), and we will denote it by $\CA\boxtimes G$. This recipe can be
placed in a very general context: given an  automorphism of a
non-commutative algebra $\CA$, and for $G$ discrete the algebra
$\CA\boxtimes G$ has the same properties as stated above.

This algebra $\CA\boxtimes G$ has various important properties. First,
the center of the new algebra is the ring of invariants of $\CA$ under
the action of $G$, and thus the center of the algebra captures the
commutative geometric quotient. In \cite{BJL, BJL2, BL} it was argued
that the center of the algebra is a natural candidate for the space
where closed strings propagate. The non-commutative geometry is
associated to D-branes, and the commutative geometry of the center is
associated to closed strings, in a spirit that follows \cite{SW}.

Secondly, the K-theory of this algebra corresponds to the (twisted) $G$
equivariant K-theory of the algebra $\CA$ \cite{Blackadar}, and thus the
topological classes of  D-branes are captured automatically, as they
correspond to K-theory classes \cite{MM, WittenK} (See also Ref.
\cite{AP}, where an attempt was made to make this equivariant
construction precise by using commutative geometry alone.) The formalism
thus allows for a computation of the K-theory classes which are
supported at singularities. These will be local quivers for each
singularity.

\subsection{ A first example: $\BC^2/\BZ_n$}

Probably the best understood orbifold in string theory is $\BC^2/\BZ_n$.
We will show how to recover the quiver diagram 
for the above just by studying the crossed product algebra.
Thus, consider two complex generators $x, y$ for $\BC^2$ which commute
with each other.
The crossed product algebra will be generated by $x, y, \sigma$, where
$\sigma$
is the non-trivial generator of $\BZ_n$ which satisfies
\begin{eqnarray}
\sigma x \sigma^{-1} &=& \omega x\\
\sigma y \sigma^{-1} &=& \omega^{-1} y \\
\sigma^n = 1
\end{eqnarray}
with $\omega^n=1$.
Consider first the center of the algebra above. Any element of the
algebra can be written in
the form
\begin{equation}
\sum_{i=0}^{n-1} \pi_i(x,y) \sigma^i
\end{equation}
The elements of the center are obtained as follows: requiring that this
commutes with $x,y$ forces us to consider only those which are of the
form $P(x,y)$ with no dependence on $\sigma$. By commuting $P(x,y)$ with
$\sigma$ we see that the monomials of $P(x,y)$ must be such that they
are invariant under the action of the group, so that we recover the ring
of invariants. The center of the algebra is generated by $u = x^n, v =
y^n, z= xy$. These together satisfy the constraint $uv = z^n$ which
identifies the algebra as the $\BC^2/\BZ_n$ orbifold.

Given the non-commutative algebra, it can be seen that the
non-commutative points are fibered over the algebraic geometry of the
center \cite{BJL}, and given a point of this commutative geometry it is
necessary to understand this fibration.

Now, we will build the irreducible representations of the algebra for a
regular point (a point which is not the fixed point). This will
correspond to having specific values of $(u,v)$. For each, there will be
a unique irreducible representation, which acts on a vector space with
basis $|0\rangle,\ldots,|n-1\rangle$. On this basis, we have
$\sigma|k\rangle = \omega^k|k\rangle$, and
 \begin{equation}
\sigma = \diag(1,\omega,\omega^2,\dots) ,\ \ \
 x = \alpha Q,\ \ \  y = \beta Q^{-1}
\end{equation}
with $Q$ a shift operator on the basis $Q|k\rangle =
|k+1\mod(n)\rangle$. This is exactly the regular representation of this
group, which we will call $R(\alpha, \beta)$. The representation space
on which the algebra is acting is a left module of the algebra
$\BC^2/\BZ_n$. $(\alpha,\beta)$ refer to coordinates in $\BC^2$. Notice
that the irreducible representations $R(\alpha, \beta), R(\omega\alpha,
\omega^{-1}\beta)$ are related to each other by conjugation by $\sigma$.
Thus the parameter space for $\alpha, \beta$ is also $\BC^2/\BZ_n$. This
is a bulk point-like D-brane. The algebra of a bulk point (or a small
neighborhood of it) is the algebra of $n\times n$ matrices tensored with
the local commutative algebra of the point, $\BC
(\alpha-\alpha_0,\beta-\beta_0)\otimes M_n$, and this is locally
Morita-equivalent to a standard commutative geometry. This is depicted
in Figure \ref{fig:orbipatch.eps}.

\myfig{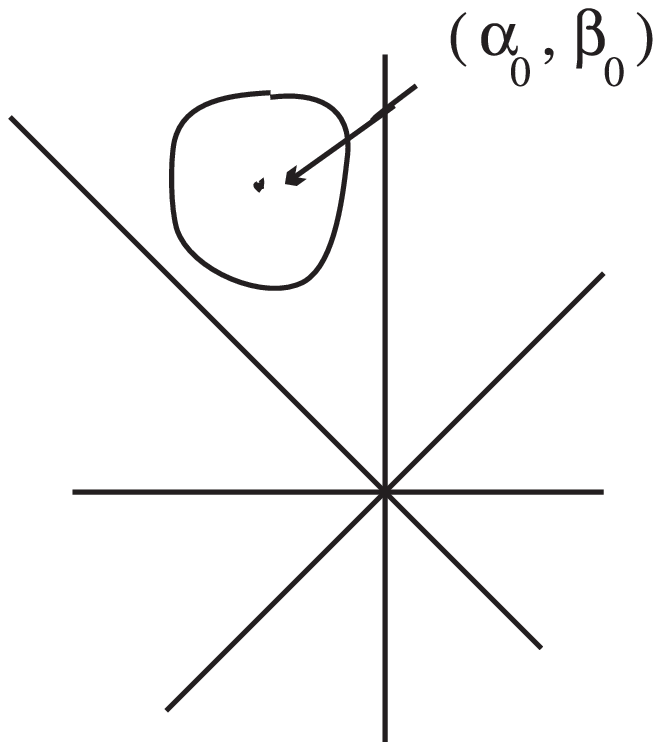}{4}{Small coordinate patch around
$\alpha_0,\beta_0$}

The only singularity in the commutative space happens when we take
$\alpha,\beta\to 0$. The representation theory of the non-commutative
algebra becomes reducible at that point, and we obtain $n$ distinct
irreducible representations labeled by $k$, namely the one dimensional
spaces $|k\rangle$  where $\sigma$ takes all of its possible different
characters. These $n$ irreducible representations of $\BZ_n$ correspond
to the brane fractionation at the singularity, and they are the
wrapped branes on the singular cycles \cite{DDG,BCD}, (these are the
new equivariant compact K-theory classes).

We can now draw a quiver for fractionation. For each irreducible
representation of the algebra  $\CA\boxtimes G$ at the singularity we
write a node. In the above example we get $n$ nodes, as we have $n$
different irreducibles at $x=y =0$.

Now, these are obtained by taking $\alpha, \beta \to 0$. The matrices of
$x, y$ are off-diagonal in a precise fashion when we take $\sigma$ to be
diagonal (they are given by the shift operator), and for each non-zero
entry of each of the $x, y$ matrices we write one arrow between the
nodes (each node corresponds to a diagonal entry in the matrix) that it
relates by matrix multiplication, and it is giving us a massless field
on the field theory of the D-brane. Later we will redo this calculation
in a more formal setting in section \ref{sec:inter} where the above will
be made precise. The resulting quiver is given in figure
\ref{fig:quiver.eps}

\myfig{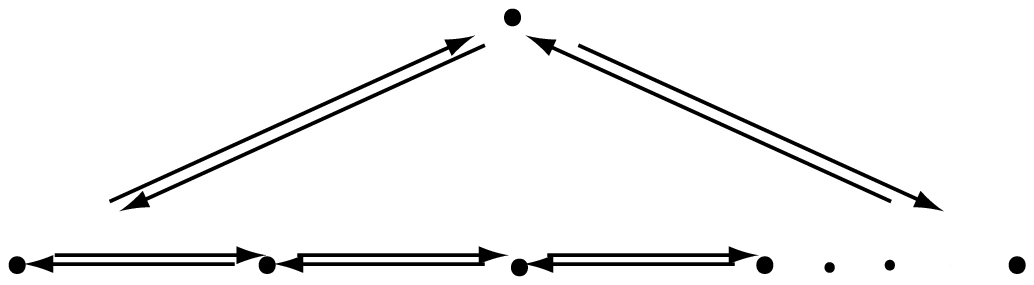}{6}{Quiver diagram for the $A_{n-1}$ singularity}

The $n$ irreducible representations associated to the fractional branes
correspond to the irreducible representations of the fixed point group
algebra (this is the group that leaves the fixed point invariant), and
these are the  irreducible representations of $\BZ_n$. Thus the nodes
are naturally associated with irreducible representations of $\BZ_n$, as
given by the quiver construction in \cite{DM}. These in turn are
associated each with a projector in the group algebra of $\BZ_n$ (a
K-theory class in the group algebra of $\BZ_n$).

The chiral fields corresponding to the individual arrows in the diagram
are then obtained by taking $\phi^1_{k,k+1} = \pi_{k} x = x \pi_{k+1}$,
and $\phi^2_{k+1,k} = \pi_{k+1} y = y \pi_k$,  with the $\pi_i$ the $n$
projectors in the group algebra. These are the independent fields on the
D-brane.

It is easy to show that the identities that the $\phi^i_{jk}$ satisfy
are exactly the ones that correspond to the equations of motion derived
from the superpotential of the quiver diagram. The projectors are part
of the algebra and they indicate on which of the nodes the chiral
multiplets begin. These are also discrete variables that are available
in the quiver diagram. Actually, the $n$ projectors are given by the
character formula $\pi_{i} = (1/n) \sum_i \omega^{k i} \sigma^i$, so one
can reconstruct the group algebra by taking linear combinations of the
projectors. Thus in this case the algebra of massless fields is exactly
the crossed product algebra and not just a subalgebra as hinted in
\cite{MarM}.

We have ignored so far the presence of D-terms. It is known that taking
the D-terms away from zero resolves the singularity \cite{DGM}, and in
the representation theory one finds that in the full $\BC^*$ algebra the
branes do not fractionate once this is done, but then one is also forced
to consider patches of algebras to account for the algebraic geometry of
the exceptional divisor. In principle the holomorphic data of both
situations is the same.

\subsection{Unorbifolding the orbifold}\label{sec:unorb}

Orbifolds by abelian groups can be undone in string theory. This
requires using the quantum symmetry of the orbifold to undo the original
twisting, and the new twisted sectors will recover what was lost
originally. Here, we will show that this operation is available in the
crossed product construction. There are two parts to this construction.
We will first analyze the previous example $\BC^2/\BZ_n$, and show how
the unorbifold recovers the original space. Then we will comment on the
general case.

The quantum symmetry on the $\BC^2/\BZ_n$ acts by phases on the twisted
sectors. Twisted sectors can only couple to the fractional branes, and
hence, the quantum symmetry exchanges the $n$ irreducible
representations of $\BZ_n$. It must also act as an automorphism of the
algebra, and it is seen that the generator of the symmetry acts by
sending $\sigma\to \omega \sigma$, while it keeps the original algebra
of $\BC^2$ unchanged. This is just as well, as the quantum symmetry is
not a symmetry of the original algebra. In the same spirit as before, we
will implement the orbifold by the quantum symmetry
$(\BC^2/\BZ_n)/\BZ_{nq}$ by introducing a new generator $\tilde\sigma$,
such that
\begin{equation}
\tilde\sigma \sigma \tilde\sigma^{-1} = \omega\sigma
\end{equation}
and $\tilde\sigma^n = 1$.

The center of the new algebra $(\BC^2/\BZ_n)/\BZ_{nq}$
is generated by
\begin{equation}
 \tilde y = y\tilde \sigma^{-1},\quad \tilde x = x \tilde
\sigma.\label{eq:chavar}
\end{equation}
Notice that these two are unconstrained, thus they represent the
geometry of $\BC^2$. Changing variables to $\tilde y, \tilde x, \sigma,
\tilde\sigma$ we get that the algebra is  a tensor product of algebras
$(\CA/\BZ_n)/\BZ_{nq} \sim \CA(\tilde x, \tilde y) \otimes A(\sigma,
\tilde\sigma)$

The factored algebra $A(\sigma, \tilde\sigma)$ has a unique irreducible
representation and as an algebra it is isomorphic to the set of $n\times
n$ matrices. Thus the algebras of $(\CA/\BZ_n)/\BZ_{nq}$ and $\BC^2$ are
related by Morita equivalence.

Notice that the extra matrices that we have obtained are discrete
degrees of freedom and they do not contribute new branches to the moduli
space-- they don't contribute degrees of freedom that are light, and the
effective actions of the light degrees of freedom for the two algebras
agree. Thus both algebras represent the same low energy physical system.

For a general abelian orbifold of a commutative space one can do the
same construction: each of the generators $y_i$ of the algebra can be
chosen to transform in a specific irreducible representation associated
to a character $\chi_i$ of $G$. The quantum symmetry will act via the
dual group $\hat G$, the group of characters of $G$. Thus we will have
generators associated to each character of $g$ which will satisfy
\begin{equation}
\chi\cdot g \cdot\chi^{-1} = \chi(g) g
\end{equation}
where $\chi(g)$ are the evaluation of the character $\chi$ in the
element $g$, and these are just numbers.

One can define new variables $\tilde y_i = y_i \chi_i$, and these are
constructed so that they commute with the elements of $G, \hat G$. Any
relation which the $y_i$ satisfy is also satisfied by the $\tilde y_i$, 
a consequence of the fact that $G$ acts by an automorphism of the
algebra of the $y_i$. Again we find a tensor product factorization where
we obtain the original algebra and the algebra generated by the $g,
\chi$. The latter algebra has a unique irreducible representation and is
thus isomorphic to the set of $|G|\times |G|$ matrices. Again the
unorbifold and the original algebra differ only by the tensor product
with $|G|\times |G|$ matrices  and are Morita equivalent. The moduli
spaces of vacua of the two agree, and the geometry associated to the two
is the same.

Notice that in the example above the fixed point set of the action of
the quantum symmetry is everything but the singular set. In this sense
in the commutative algebra of the center the quantum symmetry is a
non-geometric symmetry. In the non-commutative geometry the action is
geometrical and it exchanges the two fractional points. These combine
into a single irreducible, while the fixed points split in two, just the
opposite process of the original orbifold which combined sets of two
points into one. This is an example of Takai duality, (see
\cite{Blackadar}, theorem 10.5.2), this fact was pointed out in
\cite{MarM} without example.

\section{Orbifold of an Orbifold and discrete torsion}\label{sec:orborb}

In this section we will develop ideas presented in \cite{BJL3}, where it
was shown that one could exploit sequences of groups $H\to G\to G/H$ to
get quiver diagrams for various orbifolds, by taking the quotient
stepwise
\begin{equation}
\CM/G \sim (\CM/H)/(G/H)
\end{equation}
We will do this exercise to point out some of the pitfalls one might
encounter in doing orbifolds of more complicated algebras, while we know
what the final product should be. A discrete group will act on the
algebra as an algebra automorphism, and thus it will be a subgroup of
$Aut(\CA)$. Of these automorphisms there are some which are trivial, as
they are  constructed by conjugation by an invertible element of the
algebra.  These do nothing to the irreducible representations of the
algebra, and they have no geometric action on the non-commutative moduli
space. Moreover, the representation theory of an algebra is well defined
only up to conjugation, and this conjugation is exactly the gauge
symmetry on the D-branes \cite{BJL}. These are inner automorphisms,
corresponding to gauge transformations and should not be considered.

The group of inner automorphisms, $Inn(\CA)$, is a normal subgroup of
the group of automorphisms of the algebra. As such, we can take the
quotient group and identify
\begin{equation}
Out(\CA) = Aut(\CA)/Inn(\CA),
\end{equation}
the group of outer automorphisms of the algebra, which is the set of
symmetries which are not gauged. Thus we should be modding out only by
elements of $Out(\CA)$.

In the case where $\CA$ is a commutative algebra, the inner
automorphisms are trivial so we get that $Out(\CA) = Aut(\CA)$.
However, the outer automorphism on a non-commutative algebra $\CA$ is
only well defined in $Aut(\CA)$ up to conjugation by an invertible
element, thus when we write the physical quotient we want, we need to
take this into account to get the new orbifold algebra right. This is a
lifting problem from $Out(\CA)$ to $Aut(\CA)$. In the example in Section
\ref{sec:unorb} the lifting was such that each element of $Out(\CA)$ was
represented by an unique element in $Aut(\CA)$, thus the lifting problem
there was trivial.

As a first step, we will consider examples where this is not an issue,
and then the crossed product construction will give all of the
information of the new algebra. Consider the orbifold
$\BC^3/(\BZ_n\times\BZ_n)$, where the group acts by
\begin{eqnarray}
e_1: (x,y,z) &\to& (\omega x, y, \omega^{-1} z)\\
e_2: (x,y,z) &\to& (x, \omega y, \omega^{-1} z)
\end{eqnarray}
with $\omega^n=1$ and we want to take the orbifold stepwise as
$(\BC^3/\BZ_n)/\BZ_n$.

At the first step, we introduce the generator $e_1$ which is
such that
\begin{equation}
e_1 x e_1^{-1} = \omega x, \ \ \ e_1 y e_1^{-1} = y, \ \ \ e_1 z
e_1^{-1} = \omega^{-1} z
\end{equation}
and $e_1^n = 1$.
Because of the orbifold, we now have a quantum symmetry that acts on the
generator of the group by multiplication by a character $e_1 \to
\chi(e_1) e_1$. This quantum symmetry is also $\BZ_n$, thus we have a
group $\BZ_n\times \BZ_n$ of automorphisms of the algebra-- one $\BZ_n$
coming from the geometric realization of the second $\BZ_n$ on the
original space, and another $\BZ_n$ from the quantum symmetry. When we
do the second $\BZ_n$ geometric orbifold, we can choose to twist it by
an element of the quantum symmetry.

Thus at the second step we introduce a generator $e_2$ and we can have
\begin{equation}
e_2 e_1 e_2^{-1} = e_1 \chi(e_1)
\end{equation}
Here, we have $n$ choices for $\chi(e_1) = \omega^{\alpha}$. This choice
in the orbifold is the prototypical example of a discrete torsion phase
for the group $\BZ_n\times\BZ_n$. For a general group $G$, the phase
will be an element of $H^2(G,U(1))$, and each such choice represents a
twisted group multiplication law.

Because of the twist by the quantum symmetry, the second $\BZ_n$ will
act on the fractional branes and permute them, thus it will identify
nodes in the quiver diagram and the new quiver will be different. This
twist is also responsible for having monodromies of the fractional
branes around the codimension $3$ singularities. We will refer the
reader to \cite{BL,BJL3} for more details.

Now let us specialize to the case where $\chi(e_1)$ is a primitive
$n$-th root of unity, so we are in the situation with maximal discrete
torsion. This was the example which was used to develop the
non-commutative moduli space theory in \cite{BJL}. There the algebra of
the three superfields was found to satisfy
\begin{equation}
[\phi_i, \phi_{i+1}]_q  = 0
\end{equation}
where $[A,B]_q = AB-qBA$ with $q^n=1$.

In our case the algebra is $[x,y]=[x,z]=[z,y]=0$ plus the relations
derived from the group algebra, and the two algebras look very
dissimilar. As $e_2$ and $e_1$ no longer commute, it is useful to change
variables (in a way similar to (\ref{eq:chavar})) to $\tilde x, \tilde
y, \tilde z$, such that $e_2, e_1$ commute with the new variables. If
this can be done, the algebra factorizes.

The appropriate change of variables is the following:
\begin{equation}
\tilde x=x e_2^{-k},\ \ \
\tilde y = y e_1^k,\ \ \ \tilde z=z e_1^{-k} e_2^k,
\end{equation}
the value of $k$ chosen so that $\omega\chi(e_1)^k = 1$.
As a result, the algebra is a tensor product
\begin{equation}
\CA/(\BZ_{n}\times\BZ_n)_{d.t.}= \CA[\tilde x, \tilde y, \tilde
z]\otimes\CA[e_1,e_2]
\end{equation}
The algebra $\CA[\tilde x, \tilde y, \tilde z]$ is non-commutative, and
it is simple to show that there are relations such as
\begin{equation}
\tilde x \tilde y = \omega^{-k} \tilde y \tilde x
\end{equation}
This can also be written $[\tilde x,\tilde y]_q=0$ with $q=
\omega^{-k}$, which is also an $n$-th root of unity. The group algebra
generated by $\CA[e_1, e_2]$ with the maximal twisting has only one
irreducible representation, and it is isomorphic as an algebra to the
set of $n\times n$ matrices. Thus the two descriptions of the algebra
$\CA/(\BZ_{n}\times\BZ_n)_{d.t.}$ that we have given are Morita
equivalent.

In the case of more general nonabelian orbifolds, this is also true
although it is much harder to show. Thus the algebra of light fields on
a regular D-brane at an orbifold singularity is Morita equivalent to the
algebra of the crossed product algebra,  which represents the quotient.
In particular,  Morita equivalence implies that the K-theory of both
algebras is the same. In different situations one representation might
be more useful than the other one, but the physical content of both is
the same.

\subsection{Lifting issues}

\subsubsection{$\BC_2/\BZ_4$}
Consider now the sequence $\BZ_2\to \BZ_4\to\BZ_2$. We will use this
sequence to calculate the orbifold $\BC^2/\BZ_4$, by going through the
long route $(\BC^2/\BZ_2)/\BZ_2$. This will serve to illustrate a few
points on the lifting issues from $Out(\CA)$ to $Aut(\CA)$. The first
step in the process has already been discussed in section
\ref{sec:crossed}. The second step is the second $\BZ_2$ automorphism of
this algebra.

We will take the automorphism to be given by the expected geometric
action
\begin{equation}
e_2: (x,y,\sigma) \to (ix, -iy,\sigma).
\end{equation}
At first sight it looks as if $e_2$ is an outer automorphism of order
four, and hence we are not dividing by a $\BZ_2$, but we must note that
\begin{equation}
e_2^2:(x,y,\sigma)\to (-x, -y, \sigma)
\end{equation}
which is identical to conjugation by $\sigma$. The automorphism is of
order $4$ in $Aut(\BC^2/\BZ_2)$, but it is of order $2$ in
$Out(\BC^2/\BZ_2)$. To obtain the proper orbifold, we do the crossed
product as before, but we need to introduce the extra relation
\begin{equation}
e_2^2 = \pm\sigma\label{eq:relations}
\end{equation}
(the choice of sign is inconsequential.) That is, we equate the trivial
outer automorphisms with an appropriate inner automorphism. The
resulting algebra is indeed that of the $\BC^2/\BZ_4$ orbifold.

Thus given a subgroup of $Out(\CA)$, we want to lift to a subgroup of
$Aut(\CA)$, and as we have seen, this might not always be trivial. If we
consider a generating set of group elements for $Out(\CA)$, we choose
one lift in $Aut(\CA)$, and we might obtain a larger group. We take the
crossed product of this new group in $Aut(\CA)$ with $\CA$, and thus
obtain a space which is orbifolded too much, in the sense that we are
gauging again transformations which were already gauged. To repair this,
we must find extra relations (as in (\ref{eq:relations})) needed to
remove the overcounting of inner automorphisms. Finding such relations
can be a non-trivial task. To see this, let us consider one such example
with a non-abelian orbifold.

\subsubsection{$\BC^2/\hat E_8$}

The orbifolds of $\BC^2/\Gamma$ which are Calabi-Yau are classified by
the ADE groups. Consider $\BC^2/\hat E_8$, where $\hat E_8$ is the
discrete subgroup of $SU(2)$ corresponding to the icosahedron group. The
group $\hat E_8$ fits into an exact sequence
\begin{equation}
\BZ_2 \to \hat E_8 \to E_8
\end{equation}
where $E_8$ is the image of $\hat E_8$ in $SO(3)$. Thus we can consider
\begin{equation}
\BC^2/\hat E_8 \simeq (\BC^2/\BZ_2)/E_8
\end{equation}
The lift of $E_8$ to $\hat E_8$ is non-trivial, as it corresponds to a
non-trivial central extension of $E_8$. In particular, this extension is
associated to the possibility of discrete torsion in $E_8$ \cite{FHHP}.

Obviously we know what final algebra we want, as it is given by the
crossed product of the algebra of $\BC^2$ and $\hat E_8$. One can easily
identify the lift of $E_8$ to $\hat E_8$. Although the action of $E_8$
on the algebra $\BC^2/\BZ_2$ is linear in the coordinates, the
coordinates do not transform as any irreducible representation of the
$E_8$ algebra. Indeed, only the elements of the center of $\BC^2/\BZ_2$
(which are gauge invariant variables) transform in this manner.

The $E_8$ leaves  the fractional brane representations at the fixed
point of $\BC^2/\BZ_2$ fixed, and on one of them it acts as the algebra
of $E_8$, while in the other one the action is twisted by the discrete
torsion cocycle. Thus the action of $E_8$ on the different fractional
branes can act with or without discrete torsion, and when one assembles
the quiver diagram one gets both possible types of representations. The
quiver is drawn in figure \ref{fig:E8.eps}. \myfig{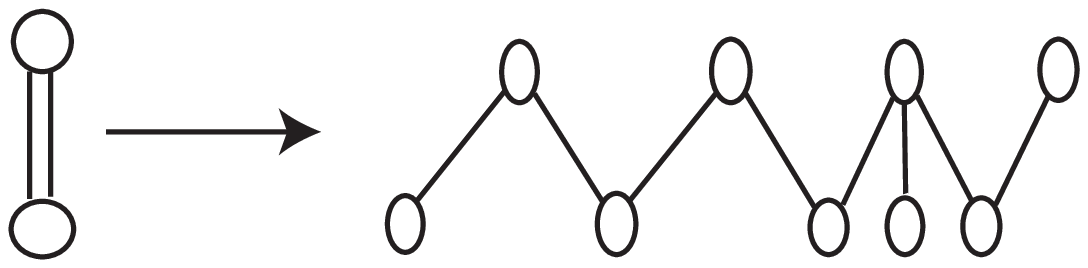}{6}{Quiver
diagram of the $\hat E_8$ singularity drawn using the two step technique
of \cite{BJL3}}

In other cases in the literature the possibility of discrete torsion
gave disconnected diagrams \cite{FHHP2}. This is because the coordintaes
transformed as representations of the group $G$, not just of the
covering group.   It has been argued that all choices of discrete
torsion can appear on the same footing\cite{Gab,Gab2}, based on a
T-duality between the orbifold $T^6/\BZ_2\times\BZ_2$ with and without
discrete torsion. Here we show an example which involves only one
non-commutative geometry without changing between `mirror' geometries.

\subsubsection{$\BC^3/\BZ_4$}

A second example we will consider is the $\BC^3/\BZ_4$ orbifold, which
has been addressed  by Joyce\cite{Joyce}, who argued  that it should
admit distinct resolutions. The orbifold acts by
\begin{equation}
(x,y,z)\to (ix,iy,-z).
\end{equation}
We will again consider the same two step procedure discussed previously
\begin{equation}
\BC^3/\BZ_4 \simeq (\BC^3/\BZ_2)/\BZ_2.
\end{equation}
However, we will twist the second $\BZ_2$ action by the quantum symmetry
of the first $\BZ_2$ (as we are instructed to do by \cite{Joyce}). In
string theory this is apparently an allowed process, but there is a
question as to how this is done for D-branes using methods of
non-commutative geometry.

The first step will give us the $\BC^2/\BZ_2\times \BC$ space, which has
generators $x,y,\sigma, z$. Now, we want the second $\BZ_2$ group to act
by
\begin{equation}
e_2: (x,y,\sigma,z) \to (ix, iy, -\sigma, z)
\end{equation}
{}From this, we obtain that in the crossed product we should have the
relation
\begin{equation}
e_2 \sigma e_2^{-1} = -\sigma\label{eq:groupnc}
\end{equation}
and one can also see that $e_2^2$ corresponds to conjugation by
$\sigma$. However, we cannot set $e_2^2 = \pm \sigma$, as this is
incompatible with (\ref{eq:groupnc}). This obstruction happens because
on a non-commutative geometry the inner automorphisms act by
conjugation, and this means that it is an element of
$GL(1,\CA)/U(1,\ZA)$. That is, the elements of $Inn(\CA)$ are invertible
elements in $\CA$ modded out by their phase which can possibly vary
continuously with the parameters of the center. Thus there is a possible
ambiguity in the lift of $e^2_2$ to an element of $GL(1,\CA)$ in order
to get the relations in the algebra right. As we see, this is a
non-trivial problem. If we orbifold by $e_2$ as it stands, it is an
automorphism of order four as we cannot find the new relations in the
algebra: we are orbifolding too much and we do not recover Joyce's
orbifold.

It is possible that there is some Morita equivalent construction
(obtained by tensoring in matrices) that gets around this problem. If
this cannot be done, then it represents a stringy obstruction to Joyce's
proposal.

\section{Non-orbifold singularities}\label{sec:conif}

So far we have seen how non-commutative geometry describes orbifold
singularities and $D$-bane fractionation at singularities. In this
section we will show that non-commutative geometries can also be used to
describe other singularities which are not of the orbifold type. As an
example, consider the conifold singularity of Klebanov and Witten
\cite{KW}. This singularity is obtained by a deformation of the
$\BC^2/\BZ_2 \times \BC$ space.

The algebra generators are $\sigma, x, y, z$, with non trivial relations
given by $\sigma^2 = 1, \sigma x\sigma = -x , \sigma y \sigma = -y,
\sigma z \sigma = z$. This is enough to give us the right quiver diagram
of the singularity $\BC^2/\BZ_2$. This fixes the commutation relations
with the discrete variable $\sigma$.

The theory has a modified superpotential, which is given by
\begin{equation}
W = \tr ( xyz - yzx) +\frac m2 \tr \sigma z^2
\end{equation}
with the presence of $\sigma$ in the last term telling us that we have
turned on a twisted sector of the orbifold (we will make this clear
later in section \ref{sec:betti}) and from here we get the additional
algebraic relations
\begin{eqnarray}
xy - yx &=& -m \sigma z  \\
yz - zy &=& 0\\
xz - zx &=& 0
\end{eqnarray}

{}From these relations we see that $z$ is in the center of the algebra,
and moreover it is redundant as a generator as we have $z=
m^{-1}\sigma[y,x]$. The choice of including or not including $z$ is
equivalent to integrating it out in the effective field theory. Also
notice that $u= x^2$ and $v= y^2$ commute with $\sigma$, and it is easy
to prove that they are part of the center of the algebra. If $m=0$, 
$xy$ is in the center of the algebra; when $m\neq 0$, this is deformed
to $w= xy+m\sigma z/2$ (equivalently, we can write this as $(xy+yx)/2$).

The relations between $u,v,z, w$ are given by
\begin{equation}
uv = w^2 - m^2 z^2/4
\end{equation}
and this geometry is the geometry of a conifold $uv = u'v'$ after a
change of variables.

Here, the slices at constant $z$ are deformations of the $\BC^2/\BZ_2$
which give a smooth ALE space. The only singularity of the conifold is
at codimension $3$ and it occurs at $u=v=w=z=0$.

To solve for the irreducible representations of the algebra we
diagonalize $\sigma$ and the elements of the center, and then $x,y$ are
seen to be off-diagonal in this basis. Away from $u=0$ we can choose
$\sigma = \sigma_3$ (the Pauli matrix), and $x=\alpha\sigma_1$. Then we
can take $y= \alpha'\sigma_1+\beta\sigma_2$, where $\alpha\beta=imz/2$
and
\begin{equation}
w = \alpha\alpha' I
\end{equation}
which as an element of the center is appropriately proportional to the
identity. Thus the representation is labeled by three complex numbers,
say $\alpha', \alpha, z$. It is straightforward to show that these are
all of the irreducible representations.

One can show that one can cover the full conifold by such patches of
representations where $u\neq 0, v\neq 0,w\neq 0$ respectively, and that
for each of those there is a unique irreducible representation. Thus
locally the space is Morita equivalent to a commutative manifold away
from the singularity. The only point which is left to discuss is $u=0,
v=0, w=0$ which is the singular point. There, the generic irreducible
representation can be constructed by the limit $z\to 0, \alpha\to 0,
\alpha'\to 0$ in the above parameterization. In this limit, the
representaton becomes reducible. Thus in this conifold the brane splits
in two when it reaches the conifold singularity. One can recover the
quiver diagram of the conifold from this splitting, and we get back the
original quiver diagram of the $N=2$ theory which was deformed. Here the
conifold is `resolved' by fractional branes. Thus we have extra
point-like K-theory classes concentrated at the origin. This is a
wrapped brane on a two-cycle, which is the cycle that one can blow-up
to resolve the conifold. Other non-commutative geometries closely
related to this system have also been analyzed with these tools in
\cite{DHOT}.

Another situation in which conifolds make their appearance in
non-commutative geometry is when we consider a deformed
$\BC^3/\BZ_2\times\BZ_2$ orbifold with discrete torsion \cite{VW,D}.

The algebra associated to this deformed orbifold is given by
\begin{eqnarray}
xy + yx &=& \zeta_1\nonumber\\
xz + zx &=& \zeta_2\label{eq:zeta}\\
zy + yz &=& \zeta_3\nonumber
\end{eqnarray}
The variables describing the center are $X= x^2,Y=  y^2,Z= z^2$ and $W=
xyz+\alpha x +\beta y +\gamma z$, with $\alpha, \beta, \gamma$
determined by the parameters $\zeta_i$. The variables satisfy
\begin{equation}
XYZ = W^2
\end{equation}
when $\zeta_i=0$, and one can show that the above deformation leads to a
space with one conifold singularity. Locally near the singularity
however, there is a unique irreducible representation of the algebra
which does not split at the commutative singular point. That is, there
is no fractional brane at the conifold point.

We conclude that the same commutative singularity can have various
non-commutative realizations, hence it is appropriate to call each of
this realizations a non-commutative resolution of the singularity. The
distinction arises because of $B$-field torsion. This is very peculiar,
but it explains why these singularities cannot be blown up in the
$T^6/\BZ_2\times\BZ_2$ orbifold with discrete torsion, as first shown in
\cite{VW}. There is no K-theory charge which can couple to the blow-up
mode, and hence there is no blow-up mode, as for every RR string field
there are D-brane sources that couple to it.

This space also has the peculiar property that there is a special place
where the representations become reducible and one dimensional (to
construct this, simply require that $x, y, z$ commute. Eqs.
(\ref{eq:zeta}) then give fixed values to $x,y,z$). This occurs at a
smooth point of the commutative algebraic geometry, and this is where
the fractional branes sit. In the undeformed theory the fractional
branes roam in codimension two singularities, and as we deformed the
theory they should remain, as K-theory should be a homotopy invariant.
As we turn on the deformation, the fractional branes that were at the
orbifold point move off to a regular point, leaving behind the conifold
singularity.

\section{ Regularity, resolutions
and intersection theory}\label{sec:inter}

String theory compactified on orbifolds can be considered a smooth
theory, as the CFT associated to an orbifold is non-singular
\cite{DHVW,DHVW2}, and moreover one can compute the spectrum and
correlation functions of the theory completely in certain situations.

We would like to see that in the non-commutative sense, an orbifold is a
smooth space.\footnote{One should compare this with the approach of E.
Sharpe via stacks \cite{S,Shaqs}. Essentially both approaches give the
same type of information\cite{Bryl}, but we believe our approach is more
convenient for computations.} Obviously, we still have singularities and
$D$-brane fractionation at singularities as viewed from a commutative
geometry point of view, as our examples have shown so far. We want to
understand how the non-commutative geometry `resolves' the singularities
and what definition of smoothness is appropriate. This should be a local
question, and moreover the non-commutative geometry is derived from
D-branes, so this question should have a D-brane answer.

Essentially what we envision here is an extension of Douglas' ideas
\cite{Doucat} to singular spaces, and non-commutative geometries. The
essential result is that one must take care in extending the formalism
to non-commutative algebras in that we must distinguish branes as left
modules, etc. A nice notion of smoothness will emerge, that of {\it
regularity} of the algebra.

The first thing we need to do then, is give a formulation of (extended)
D-branes. For the time being, we will assume that the D-branes are BPS
and can be realized holomorphically. A D-brane should be associated with
a K-theory class, and  this means a formal difference of two vector
bundles. On a non-singular commutative space a vector bundle is a
locally free module over the algebra of functions (that is, it has
constant rank). Thus a holomorphic vector bundle on a complex manifold
can be represented by a locally free sheaf globally.

Since not all D-branes are space filling, it is natural to replace
locally free sheaf by a coherent sheaf; also when one considers branes
embedded into branes this appears naturally \cite{HM}. A coherent sheaf
$S$ is locally the cokernel of a map between two locally free modules
\begin{equation}
\CA^n\to \CA^m \to S \to 0
\end{equation}
where the above is an exact sequence. If the complex dimension of the
manifold is $d$ (and the manifold is non-singular) any coherent sheaf
admits (locally) a free resolution of length $d+1$\cite{Hart}. That is,
we can find holomorphic vector bundles (locally free sheaves)
$F^0,\dots, F^d$, such that there is an exact sequence  of sheaves
\begin{equation}
0\to F^d \to F^{d-1}\to \dots \to F^0\to S \to 0
\end{equation}
With this construction we can write the $K$ theory class of $S$ as
\begin{equation}
K(S) = K(F^0) - K(F^1) + K(F^2) - K(F^3) +\dots+(-1)^d K( F^d)
\end{equation}
and in principle we should be able to replace $S$ by it's resolution.
When this is done we are starting to see the appearance of homological
algebra, which is the study of the topological properties of the
complexes of objects in the category of sheaves. To include both branes
and antibranes it is more convenient then to introduce the derived
category of coherent sheaves \cite{Doucat}. This category is built out
of complexes of coherent sheaves with some identifications, see
\cite{Weibel,GManin} for more details. Recent work on this subject has
also appeared in \cite{AL,HelMc,GJ}, but they focus on either
commutative geometry tools or linear sigma models for these problems.

In the non-commutative setting we want to do the same thing. There are a
few differences however. We will still define a coherent sheaf as a
cokernel of a map between two locally free left modules, the map being a
left module map
\begin{equation}
\CA^n \to \CA^m \to S\to 0
\end{equation}
The $K$-theory we are after is the standard algebraic K-theory, defined
by the projective left modules of the algebra \cite{Ros}. The above map
should be considered as a map of left modules.

When we are studying local geometries which are just $n\times n$
matrices over the local commutative ring the above consoderation is
enough. For infinite matrices we have to be more careful and we have to
replace the above by
\begin{equation}
P^1\to P^0 \to S \to 0
\end{equation}
where the $P^i$ are projective of finite type. The two definitions are
identical for the $n\times n$ matrices, but the second is better suited
for the infinite dimensional case. The ring as a left module is of
infinite rank in the case of infinite matrices.

Now, we have coherent sheaves of left modules\footnote{Originally it was
suggested to use bimodules in \cite{BJL} because the representation
theory themselves are bimodules, but not the space on which they act on
the left. A bimodule is more appropriate to study strings between two
D-branes, hence the confusion.} of the ring as our candidates for
extended D-branes. The next thing we need is a definition of smoothness.
The proper definition is related to the existence of projective
resolutions. A non-commutative ring is defined to be regular at a point
$p$ if every coherent sheaf (localized at $p$) admits a projective
resolution (of finite length $d+1$), that is, we have
\begin{equation}
0\to P^d \to P^{d-1} \to \dots \to P^0 \to S \to 0
\end{equation}
and the projective left-modules are the equivalent of vector bundles
over the non-commutative space, with $d$ the complex dimension of the
manifold. Thus, we have a condition to check for our examples. This is a
good definition because the $K$-theory of the space is constructed from
projective modules. Thus in this setting every brane has a well defined
K-theory charge. This condition also seems to be required in order to be
able to construct the $K$-theory of the string background directly from
the derived category, (see \cite{Srinivas}, ch. 5 for more background).
There is a subtle technical point to be made here. On analytic
manifolds, vector bundles are both projective and injective, because
once one imposes a metric on the bundle one can take orthogonal
complements, but this is not true in the algebraic category of sheaves.
Injective resolutions always exist, and that is how one usually defines
the global sections functor \cite{Hart}. However injective sheaves are
notoriously large (they are not finitely generated) and at least in this
sense one would be forced into considering an infinite number of  brane
anti-brane pairs, see however \cite{Hori,WittenK2}. Hence, these cannot be
interpreted in terms of $K$-theory in a straightforward manner. Locally
projective (finite) resolutions do not always exist. Indeed the
commutative (singular) space of the center fails to have this property
and is therefore non-smooth in this sense as well.

When we are at a non-singular point in the commutative sense, the local
non-commutative algebra is Morita equivalent to the local commutative
algebra. The local commutative algebra is regular, so any (finitely
presented)  module over it (coherent sheaf) admits locally a projective
resolution. Because Morita equivalence establishes an isomorphism on the
categories of modules over the rings, one gets via this ismorphism a
local projective resolution for any sheaf over the non-commutative ring.
Therefore the non-commutative ring is also regular at these points. Thus
the only places where we need to check the regularity condition is
exactly at the singularities.

Here we will analyze the case $\BC^2/\BZ_n$, and we will concentrate
only on the compactly supported K-theory classes. That is, we will
consider a torsion sheaf at the origin, and all such are fractional
branes.

A single fractional brane is an irreducible representation of the
algebra, and as a complex vector space has dimension one. As
$\sigma^n=1$, there are $n$ natural candidates for projective modules
over the local algebra, and they are related to the $n$ possible
projection operators $\pi_k= \frac1n\sum \omega^{ki}\sigma^i$ on the
group algebra. These are the only $n$ non-trivial projectors at the
origin, and they cannot be made homotopic to one another, as they act by
different values on the irreducible representation at the origin. Away
from the origin, one of $x,y$ is invertible, as their $n$-th power is,
and we can use conjugation by this invertible element to turn the
different projectors into each other.

Thus, as our local left modules we take $M_i=\CA \pi_i$. We can use
these to produce a sequence for each irreducible representation of the
algebra $S_k$, as follows
\begin{equation}
0\to\CA \pi_k \to \CA \pi_{k-1}\oplus \CA \pi_{k+1}  \to \CA
\pi_k \to S_k\to 0 \label{eq:projresk}
\end{equation}
with the Koszul complex on each of the resolutions. That is, the module maps are
given by multiplication with $x,y$ on the right. This can be understood in the 
above sequence because each of the modules is a submodule of the ring $\CA$ itself, 
multiplying on the right by $x,y$ gives us new elements in $\CA$ which belong to 
different projective submodules. Moreover
multiplication on the right commutes with multiplication on the left, so these maps are 
left module morphisms. To be more explicit, if we have elements $a\in \CA\pi_k, (b_1,b_2)
\in (\CA\pi_{k-1}, \CA\pi_{k+1}$ of each of the modules the maps are given by
\begin{eqnarray}
a & \to & (ay, ax)\\
(b_1,b_2) &\to& b_1x -b_2 y
\end{eqnarray}

Thus a fractional brane admits a projective resolution. A general proof
that the ring is regular is beyond the scope of this paper (such a proof
will be presented elsewhere \cite{BG}).

Now that we have some sense in which we can call the orbifold smooth
from the non-commutative geometric point of view, we need to interpret
the module maps in the above exact sequences. For the fractional branes,
it is clear that the module maps give isomorphisms on the fibers away
from the origin. These isomorphisms of the fibers at infinity are
familiar from K-theory with compact support, and they have been
interpreted as tachyons by Harvey and Moore \cite{HM2}. In our case
notice that the tachyon profile is analytic, and at first sight this
would seem like a disaster, as the energy associated to this
configuration would be infinite. However, we are analyzing complex
vector bundles, and for these, the structure group is complexified. The
fact that the functions are divergent at infinity is a gauge artifact, 
as we are not using a hermitian connection on the bundle. If this is
done, the above construction is just a continuous profile which is
bounded at infinity. Since we have not mentioned metrics at all,
introducing a metric now makes us consider hermitian Yang-Mills bundles
instead of holomorphic vector bundles (sheaves), but the two moduli
spaces one can compute should be related. In this language one would
talk of complexes of differential algebras instead, see for example
\cite{Doucat,AL,Dia} and one would use the cohomology groups of
$\bar\partial$ twisted by a holomorphic vector bundle. These results
should be equivalent \cite{GH}.

\subsection{Counting open strings and quiver building}

Given two  $D$-branes, one wants to consider counting the number of open
massless string states between them. This counting should be invariant
under Morita equivalence, should be naturally associated to the coherent
sheaf (or to an element of the derived category which represents it),
and it should come from a homological framework.

The invariance under Morita equivalence is obviously a symmetry we have
been considering already, and we have shown that Morita invariant
algebras represent the same space, thus the topological data we can
extract from $D$-branes should be the same on each possible
representation. Obviously we want an algebraic machine with which to
calculate these, and the BRST operator of open strings\cite{Doucat,AL}
suggests that the result should be homological.

Naturality here is interpreted in the functorial language. Since we are
considering two left modules of the algebra $\CA$  and maps that
preserve this condition (the projective resolutions), the natural choice
is to use $hom_\CA(A,B)$ for two modules $A,B$. This is invariant under
Morita equivalence, as two rings are Morita equivalent if their
categories of modules are isomorphic and this includes morphisms of
modules. Since $hom$ is not an exact functor one uses instead the
derived functor of $hom$. Thus the groups which are important for our
computation are the $\Ext^i(A,B)$, which can be extended to the derived
category. Although for our computations this is all we need, in general
it seems to be important to define the category of branes by also
counting the ghost number in the open string fields \cite{Doucat, Dia,
AL}, that is, one should keep all of the information in the grading of
the $\Ext^i$. The contribution to the intersection index will depend on
$i\mod 2$ alone however.

The definition of the $\Ext$ groups can be done using the projective
resolutions of the module $A$, $\dots\to P_0\to A\to 0$. One obtains  a
complex
\begin{equation}
0\to hom(P_0,B) \to hom(P_1,B) \to \dots
\end{equation}
The $\Ext^i(A,B)$ groups are the homology groups of the above complex
(see \cite{Weibel, Osborne} for example), and they are independent of
the choice of the resolution. Thus one can calculate these explicitly if
one knows the projective resolution of an object.

The group $\Ext^1(A,B)$ is special, as it counts locally the space of
deformations
\begin{equation}
0\to A\to C \to B \to 0
\end{equation}
for gluing two branes $A,B$ to form a third $D$-brane which is a
(marginally) bound state of the two branes, $C \sim A\oplus B$. This is
exactly how one understands the process of brane fractionation to make a
bulk brane.

Let us take for example $\BC^2/\BZ_n$. We have already computed the
projective resolution of all of the fractional branes in
(\ref{eq:projresk}). Now we want to understand the intersection product.

Thus consider the $\Ext^i(S_i,S_k)$ groups. Since each of the $\CA \pi_i$
are generated by the unique section in degree zero $\pi_i$, the image of
this element generates the module map. Moreover the $S_i$ are
concentrated in degree zero, and the map reduces to a computation in the
group algebra, namely $hom(\CA \pi_l, S_k) \sim \hom_G(\chi_l, S_k)$,
where $\chi_l$ is the representation of the group $G$ in degree zero in
the module. Since each fractional brane corresponds to a unique
irreducible, the dimension of the above vector space is $1,0$ depending
on whether $l=k$ or not. In the $hom$ complex one obtains, the maps $x,y$
act by zero, thus the homology of the complex is the complex itself.

We find the following  results
 \begin{eqnarray}
 \dim\Ext^0(S_i, S_k) &=& \delta_{ik}\\
 \dim \Ext^1(S_i, S_k) &=& \delta_{i,k-1}+\delta_{i,k+1}\\
 \dim \Ext^2(S_i,S_k)&=& \delta_{ik}
 \end{eqnarray}
Notice that the quiver diagram of the singularity is captured by
$\Ext^1(A,B)$. In considering topological string states  these are the
states that correspond to the vector particles, while $\Ext^0$ counts
the possibility of tachyon condensation between branes and anti-branes.

The intersection number between two branes is the Euler characteristic
of the above complex (perhaps up to a sign). We obtain that the self
intersection of the fractional branes is $-2$, and that the matrix of
Euler characteristics is exactly the Cartan matrix of the extended
affine $A_n$ algebra. This is also the intersection form of the resolved
orbifold by blow-ups, where each fractional brane wraps one of the
exceptional divisors, plus a brane that wraps all of them with the
opposite orientation (the extended root of the system). When the cycles
are blown-down, these states are mutually $BPS$.

The advantage of formulating the problem in terms of homological algebra
is that the above procedure gives us a recipe for computing these
numbers in a general non-commutative geometry singularity which is not
of the orbifold type. We can compute these numbers  for the fractional
branes in the Klebanov-Witten conifold and we obtain the quiver diagram
consisting of two nodes with two arrows between them, which contains
only the fields which are not lifted by the mass deformation,
exactly reproducing the field theory data of the CFT.

There is another point to make which is useful. This intersection
formula between fractional branes is symmetric if the complex dimension
of the space we are orbifolding is even, and antisymmetric if the
complex dimension of the space is odd \cite{BG}. This is exactly what
one expects from the topological intersection pairing of the BPS
D-branes in the mirror manifold, which are even or odd real cycles
depending on the dimension of the Calabi-yau space.

\section{Closed string states}\label{sec:betti}

So far we have concentrated on open string states alone, with a brief
mention of the center of the algebra as the place where closed strings
propagate. The description of backgrounds we are constructing depend on
having $D$-branes on them, and can be mainly considered as open string
theories.

Open string theories will contain closed strings, so it is natural to
ask if there is a way to construct closed string states from a
non-commutative algebra. We will restrict ourselves to topology, and
hence to topological string states alone.

As argued in \cite{BL4} the natural way to introduce closed string
states is suggested by AdS/CFT duality \cite{M,GKP,W}, where the single
trace gauge invariant states of a field theory on $D$-branes are
naturally associated to closed string states in the near horizon
geometry. Here, we want to exploit this philosophy and write generic
closed string states as `gauge invariant' single trace operators.

The general construction should also accommodate the notion that closed
topological string states are harmonic forms on the space, so it is
natural to look for them as traces of non-commutative forms, and to
check that they are closed. That is, one would want the analog of 
deRham currents for a non-commutative algebra. A second point to consider
is that topological closed strings should be associated with
Chern-Simons couplings to $D$-branes.

We are mainly interested in point-like D-branes, so we will introduce
the couplings to these states as traces of algebra elements in the
representation associated to a D-brane. We will require that the values
of these traces depend only on the homotopy class (K-theory class) of a
D-brane, and as such they are topological.

It is best to work with our familiar example $\BC^2/\BZ_n$. We have $n$
distinct point-like  K-theory classes, one for each node in the quiver
diagram of the singularity. They are distinguished by the eigenvalue of
$\sigma$. Thus it is natural to take the traces
\begin{equation}
\tr(\sigma^k)
\end{equation}
as our candidate closed string states, namely the characters of the
group elements. Notice that as the quantum symmetry acts on $\sigma$,
there is an action of the quantum symmetry on the closed string states.
These are then naturally twisted sectors.

Notice that only the one corresponding to $\tr(\sigma^0)$ is untwisted.
On a bulk brane, the traces for twisted states give zero, as a bulk
brane has one of each irreducible representation of the group algebra.
Thus we have $n$ traces, one of which is untwisted. The blown-up ALE
space also has $n$ closed forms with compact support, one representing
each of the $n-1$ blown up self-dual cycles, and one more for the class
of the point.

Traces like $\tr(\sigma^k P(x,y))$ for $k\neq 0$ all vanish outside the
singularity and moreover they evaluate to zero on the fractional branes.
Thus these don't contribute more states. Other traces like $\tr P(x,y)$
depend on the location of a brane in moduli space, unless the polynomial
is of degree zero.

A generalized trace that can also measure extended D-brane charge 
should be (locally) of the form
\begin{equation}
\tr(a_0 da_1 da_2\dots da_n)
\end{equation}
for elements of the algebra, and these are exactly the elements of the
cyclic homology of the algebra \cite{Connes}. On bulk points, these will
reduce to ordinary elements of deRham cohomology on the local
commutative space, via Morita equivalence in cyclic homology.

\subsection{Betti Numbers for global non-commutative Calabi Yau spaces}

In this section we will give a prescription for computing the
Betti numbers of the orbifold $T^6/\BZ_2\times\BZ_2$ with and without
discrete torsion.  Once these two examples are dealt with, we will give
a general recipe for computing the Betti numbers of certain singular
Calabi-Yau $3$-fold with singularities in codimension $2$ with a
non-commutative resolution of singularities.

To begin with the computation of the Betti numbers of our main example,
we have to deal first with the $\BC^2/\BZ_2$ singularity, which is the
local version of the singularities of codimension two.

We already have the algebra of $\CA=\BC^2/\BZ_2$. String theory predicts
that in these singularities we generate as many blow-ups and complex
structure deformations as there are massless twisted sectors, in this
case one. Blowing up or deforming the singularity we find one square
integrable harmonic form of type $H^{1,1}$. The fractional brane in the
blow-up becomes a brane wrapped around the resolved cycle, and
integrating this (normalized) $(1,1)$ form over the compactly supported
brane can serve to count how many branes are wrapping the cycle. This is
basically the topological coupling of the D-brane data to the RR
potential in string theory, the Chern-Simons coupling. The non-torsion
part of the lattice of K-theory can be paired with cohomology with
complex coefficients, and the two carry the same information
\cite{Connes}.

Thus counting the cohomology classes can be done by computing invariants
of K-theory classes. The closed string states are then the dual space
for these objects (couplings in the action for open string states), and
one should basically think of them as integrating over cycles (D-branes
with some excited states on them) in the manifold and producing numbers
(the action for the configuration). When we write an expression in terms
of traces, this will be implicitly understood.

In our local case, we have brane fractionation at the origin, and we
should be able to compute invariants of these point-like classes that
are additive in the K-theory classes and vanish for the non-fractional
brane. Indeed, the fractional branes correspond to the one-dimensional
irreducible representations of the algebra of $\BC^2/\BZ_2$. The ring at
the singularity consists only of the group algebra, and the traces
reduce then to characters of the group elements in the associated
representations. There are only as many linearly independent traces as
there are conjugacy classes of elements in the group $G$, which is
identical to the number of irreducible representations of the group and
to the number of closed string twisted sectors.

In our case, there is the trivial trace of $1$, corresponding to the
trivial representation of the group, and this trace counts the total
number of bulk zero branes.\footnote{This coupling measures the total
D0-brane charge. On extended branes this is an integral over the whole
brane, and it essentially measures the top chern class of the fiber
bundle, with perhaps some corrections from geometry, see \cite{MM}.}
There is a second trace, $\tr (\sigma)$. For the two
one-dimensional representations at the fixed point, the value of this
trace is $\pm1$, so they wrap the singular cycle with the opposite
orientation. The state $\tr(\sigma)$ vanishes on the bulk
representation. Thus the support of the state (the points in the
non-commutative moduli space where $\tr(\sigma)\neq 0$) consists of the
two points at the origin. This is also a twisted sector, as when we act
with the quantum symmetry the state changes sign.

Since $\sigma$ can be chosen globally constant in the representation
theory, we want to argue that $d \tr(\sigma) = \tr(d\sigma) =0$, so that
$\tr(\sigma)$ is a cohomology class for $d$.\footnote{Here we are
loosely extending $d$ to the singular point.}

We also know that we can blow-up the cycle, and that this twisted class
becomes a $(1,1)$ form on the blow-up. The fractional brane should be
imagined to be wrapping a 2-sphere cycle on the orbifold \cite{DDG,BCD},
and in the trace we are integrating over this cycle. Thus our twisted
state should be identified with a $(1,1)$ class.

In terms of traces of elements of the algebra, it is not clear why these
should correspond to $(1,1)$ classes. However, their holomorphicity
suggests that they should be associated with $(k,k)$ classes for some
$k$.\cite{Ruan, Ruan1,CR}

Now, we have these classes locally on any patch of our non-commutative
space, and we need to glue them together so that we have global
holomorphic forms.

In the case of $T^6/\BZ_2\times \BZ_2$ without discrete torsion, we get
one such trace for every fixed plane. Notice that the group elements
survive the quotient (they are not eliminated by a Morita equivalence),
and on each codimension two singularity there is one group element which
does not vanish, thus for each fixed plane we have one blow-up mode.
This is also the result we get from conformal field theory\cite{VW}.
Remembering the structure of the cover of the singularity (this is
depicted in figure \ref{fig:nccover1.eps}), the new class corresponds to
$b^0(\CP^1\oplus \CP^1)-1$ (the $0$-th Betti number), where we subtract
the trivial class of the brane in the bulk which is not a new class.

\myfig{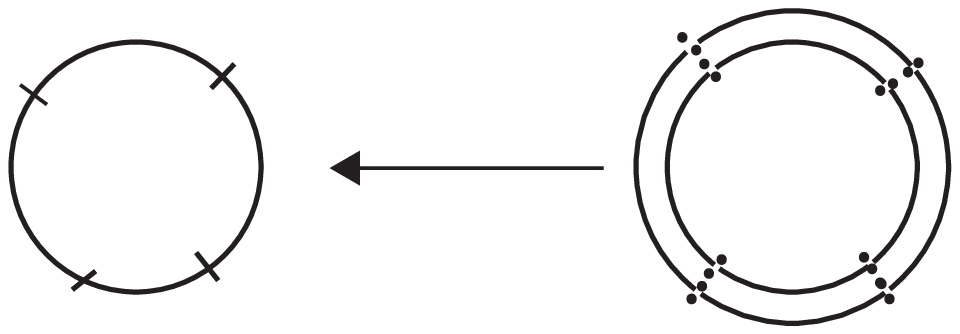}{6}{Noncommutative cover of singularity without
discrete torsion: the special points correspond to the intersection of
two curves of singularities}

For the case of the orbifold with discrete torsion, we have to remember
that we changed variables to the $\tilde y$, and that we can eliminate
the group variables by Morita equivalence.

In this case, at the singularity the traces that do not vanish are
\begin{equation} 
\tr(\tilde y_i^k) 
\end{equation} 
with $k$ odd. Now, when we move the branes on their moduli space, these
vary and there is no new $H^{1,1}$ class, as there is no global
holomorphic function which is odd in the $y$ on the singularity. This is
because our double cover of the singularity is connected and compact,
thus $b^0(T^2)-1 =0$. From string theory this is just as expected, but
now we also expect one new $H^{2,1}$ class for each cycle.

However, the double cover of the singular $\CP^1$ is a torus, and we
have a natural $1$-form on this torus which is it's abelian one-form.
This cover is depicted in figure \ref{fig:nccover2.eps}. The expression
in Weierstrass form for this one-form is
\begin{equation}
\tr dy_i/x_i = \frac {d(\tr(y_i))}{x_i}
\end{equation}
which we can also see has support on the singularity, as elsewhere
$\tr(y_i)=0$. This new one-form is wrapped on the same $(1,1)$ cycle as
we had before, so it should correspond to an element of $H^{2,1}$ of the
 Calabi-Yau manifold. Hence we get a contribution to $b^{2,1}$ from
$b^{1,0}$ of the singular set\cite{BL4}.

\myfig{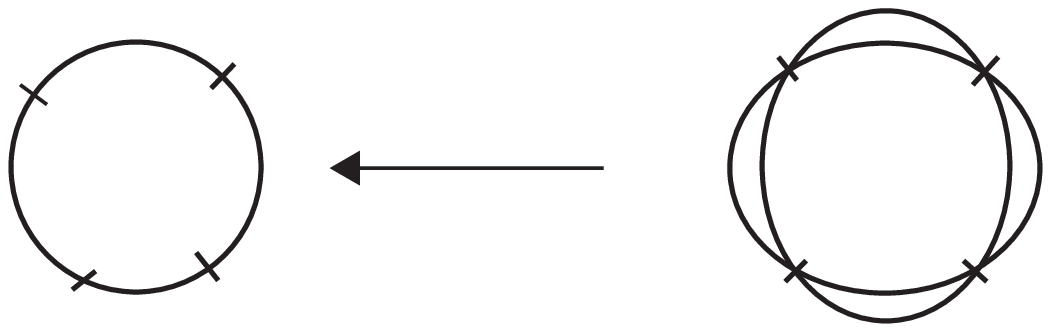}{6}{Noncommutative cover of singularity with
discrete torsion: at the special points the fractional branes have
monodromies \cite{BL}}

Notice that now, for the case without discrete torsion, the singular set
is fibered non-commutatively by two copies of $\CP^1$, each of which is
simply connected, hence there is no contribution to $H^{2,1}$, as there
are no holomorphic one-forms on the singular set. This is also in
accordance with conformal field theory calculations.

For $b^{2,2}$ of the Calabi-Yau we need to do the same analysis. By
Hodge duality, this should be equal to $b^{1,1}$ of the CY manifold. In
our case, both the $T^2$ and the $\CP^1\oplus \CP^1$ have  classes in
dimension $(1,1)$ that can contribute.

Now let us discuss more general Calabi-Yaus.
A codimension two singular curve  contributes
to $H^{1,1}$ according to the number of connected components of the
non-commutative fibration. One also sees  contributions to $H^{2,1}$ by
an amount equal to $H^{1,0}(C)-x$, where $x$ is the number of global
one-forms shared with the manifold (so if the manifold is simply
connected, $x=0$). We also contribute to $H^{2,2}$ by the same amount as
to $H^{1,1}$, mainly because we are imposing Hodge duality by hand, as
it holds on the covers of the individual singular cycles. The orbifold
cohomology theory indicates that this can be assumed in general
\cite{Ruan}, and since we are in a Calabi Yau manifold the cohomology
classes will not have fractional degree.

For codimension three orbifold singularities in a Calabi-Yau, we get
local fractional branes. These singularities contribute to both
$H^{1,1}, H^{2,2}$ of the global Calabi-Yau by a number equal to
$(H^0-1)/2$, and it is also known that they don't contribute to
$H^{2,1}$.  Usually orbifold points are at walls between distinct
phases\cite{Greene,BGLP} but all of these have the same Betti numbers
and moduli spaces of complex structure deformations, so even though we
are at a transition point between topologies the non-commutative results
we have obtained are meaningful.

\section{Outlook}

Strings propagating on singular spaces might not be singular, and here
we have made a proposal on how to interpret this fact by exploiting the
non-commutativity of open strings. The proposal has two geometries: one
non-commutative geometry of the open strings, and a derived commutative
geometry of closed strings (in our case defined by the center of the
local algebras). The main ingredient in this resolution is the use of
the category of D-branes to define the geometry. This simple setup
provides a remarkable amount of topological information, namely, one
can calculate the intersection theory of branes very generally. One can
also explicitly see the existence of closed string twisted sectors at
singularities and this is how one usually resolves singularities by
closed string fields.

However, the examples we have provided are not general enough to capture
all of the possible non-commutative geometries that strings can describe
and should be viewed as a first step towards a more general theory. For
example, the non-commutative plane is left out of our construction,
namely because the algebra is not a finite matrix algebra over the
center;  this non-commutative geometry would be a
fibration over a point.

{}From a Mirror Symmetry perspective we have only dealt with the $B$ model
physics, so one would also be interested in exploring how to realize
these ideas in the $A$ model as well, and in particular how to calculate
stringy corrections to different quantities in this setup, see for
example \cite{AV}. It would also be interesting to extend these
non-commutative ideas to the Landau Ginzburg phases of string theory.

However, given our particular realization of geometry it is also very
difficult to introduce a worldsheet string action. Should one use a
non-commutative sigma model in the spirit of matrix strings \cite{DVV}?
Or should we use instead a commutative sigma model action with a
prescription for dealing with the singularities? It has been shown that
writing matrix theories on general backgrounds is difficult\cite{DO};
however, by restricting ourselves to topological string theories we
might be better off, as the number of string fields that one needs to
consider becomes finite and the theory does not depend on the exact
properties of the metric.

The description we have given should only be applicable to situations
where the singularity can be resolved perturbatively in the string
theory. This is particularly evident because closed string states and
$D$-branes wrapping cycles are treated in a completely different
fashion. It is unclear if non-commutative geometry will play any role at
all at the conifold transition.

These and other issues are currently under investigation.

\bigskip

\noindent {\bf Acknowledgments:} It is a pleasure to thank M. Ando, M.
Artin,  P. Aspinwall, D. R.  Grayson, V. Jejjala, S. Katz, E. Martinec,
Y. Ruan, and C. Vafa for numerous discussions and comments in the course
of this work. Supported in part by U.S. Department of Energy, grant
DE-FG02-91ER40677.

\providecommand{\href}[2]{#2}\begingroup\raggedright\endgroup

\end{document}